\documentclass[10pt,letterpaper]{article} \usepackage[top=0.85in,left=1in,footskip=0.75in]{geometry}\usepackage[utf8]{inputenc}\usepackage{cite}\usepackage{nameref,hyperref}\usepackage[right]{lineno}\usepackage{microtype}\usepackage{float}\DisableLigatures[f]{encoding = *, family = * }
\usepackage{changepage}\usepackage[aboveskip=1pt,labelfont=bf,labelsep=period,singlelinecheck=off]{caption}\makeatletter\renewcommand{\@biblabel}[1]{\quad#1.}\makeatother\usepackage{lastpage,fancyhdr,graphicx}\usepackage{epstopdf}\usepackage{color}\definecolor{Gray}{gray}{.25}\usepackage{graphicx}\usepackage{sidecap}\usepackage{wrapfig}\usepackage[pscoord]{eso-pic}\usepackage[fulladjust]{marginnote}
\begin{document}\vspace*{0.35in}\begin{flushleft}

{\Large
\textbf\newline{Directional Emission From High-Q Asymmetric Hollow Whispering Gallery Resonators}
}
\newline
Amal Jose\textsuperscript{1,+},
Ramgopal Madugani\textsuperscript{1,2},
Christophe Pin\textsuperscript{1},
S\'ile {Nic Chormaic}\textsuperscript{1,*}
\\
\bigskip
\bf{1} Light-Matter Interactions for Quantum Technologies Unit, Okinawa Institute of Science and Technology Graduate University, Onna, Okinawa 904-0495, Japan\\
\bf{2} Current address: RIKEN Center for Quantum Computing (RQC), RIKEN, 2-1 Hirosawa, Wako, Saitama, 351-0198, Japan
\\
\bigskip
{$^+$} amal.jose@oist.jp, * sile.nicchormaic@oist.jp

\end{flushleft}

\begin{abstract}
Whispering gallery mode (WGM) resonators with broken rotational symmetry exhibit phenomena that are generally absent in conventional symmetric cavities, including directional emission. While such effects have been extensively investigated in solid and planar resonators, they remain largely unexplored in hollow, three-dimensional cavities, where optical confinement is fundamentally altered by the thin wall geometry. Here, we demonstrate controlled fabrication of asymmetric silica microbubble resonators through anisotropic expansion during the microbubble formation process. X-ray tomography confirms the resulting three-dimensional geometry and allows us to quantitatively characterize the cavity deformation. Despite the broken rotational symmetry, the resonators maintain whispering gallery modes with high loaded Q-factors exceeding $10^5$  for moderate deformations. Optical characterization reveals that only certain resonances exhibit  directional emission, whereas neighboring modes retain conventional isotropic behavior. Two-dimensional numerical simulations reproduce the observed emission characteristics and indicate that deformation-induced leakage of higher-order modes provides a plausible mechanism for the directional radiation. These results establish asymmetric hollow microbubble resonators as a platform for investigating light transport in three-dimensional asymmetric whispering gallery cavities while preserving the high-Q performance required for photonics applications.
\end{abstract}

\section{Introduction}
Whispering gallery modes (WGMs) are resonant eigenmodes of cavities in which light orbits along the boundary of the resonator's concave surface via total internal reflection (TIR).
For circular symmetry reasons,  the orbital angular momentum projection along the symmetry axis of the resonator is conserved along the light trajectory, resulting in resonant modes with ultrahigh Q-factors, relatively small mode volume, long photon lifetime, and isotropic radiational decay \cite{isotropic_radiation1992, highQ1998,highQ1999,highq2018,highq2023_2}. 
This type of photonic resonator is particularly suited for enhancing nonlinear phenomena \cite{Nonlinear2004,Yang:16c,Tian:22d,nonlinear2022}, lasing effects \cite{4451197,10.1063/1.3277024, Fang:17, lasing2}, enhancing light-matter interactions in cavity QED experiments \cite{highQ1998, CQED2021-2}, and for enabling ultrahigh sensitivity optical sensing \cite{GS2}. 

Breaking the rotational symmetry by introducing a smooth deformation has led to new phenomena being observed, such as directional emission \cite{directionalemission1998, Refractive-Escape, Directionalemission2020}, chaotic mode excitation \cite{chaosexcitation2010, cat2}, chaos-assisted tunneling\cite{CAT1,CAT2017}, and local chirality at boundaries with a small curvature \cite{intracavityraydynamics2012}. Although most of those phenomena have been studied in (quasi) two-dimensional resonators, e.g., microdisks\cite{IWGM1, IWGM2, Rodemund1}, very few works have looked at asymmetric three-dimensional microcavities, e.g., microspheres, despite some promising results \cite{chaosresult2010,chaos3d2025}. One reason for the limited experimental studies on 3D devices is the difficulty in controlling deformation parameters during fabrication, making it challenging to obtain reproducible results and conduct parametric studies. In contrast, microrings and other waveguide-based 2D resonators can be much more robust against deformations, allowing, for example, racetrack or spiral geometries to still exhibit high Q-factors \cite{spiral2004,  Billiard2024}.

Thin-walled microbubble resonators \cite{watkinsbubble, microbubblefab}, i.e., hollow three-dimensional WGM resonators with a wall thickness on the order of a wavelength,  can be viewed as having properties somewhere between those of a ring and a solid sphere resonator. They offer 
radial confinement due to the thin wall, while still relying on curvature-induced confinement in the polar direction. It is, therefore, not straightforward to predict the effect of asymmetry in such a photonic device.   However, unlike solid 3D asymmetric resonators, where chaos-assisted light transport can follow a free trajectory within the resonator, the hollow structure of microbubbles restricts radial trajectories, thereby forcing them to be predominantly on the boundary if they exist.
In this work, we develop a  technique for fabricating deformed microbubble resonators with well-controlled asymmetry. We characterize the 3D geometry of the  microbubbles using X-ray tomography \cite{chaos3d2025} and  spectrally characterize their whispering gallery modes.  We observe deformation-induced directional emission and, via numerical simulations, present a plausible mechanism for this phenomenon. Such a deformed resonator may also provide insight into chaos-assisted light transport in radially restricted systems.
 
\section{\label{fab}Fabrication Method}
To fabricate the deformed microbubbles with good control over the asymmetry and reproducibility, we implemented a three-step fabrication method. The first two steps followed the previously reported method for fabricating symmetric microbubbles \cite{watkinsbubble}, whereas the third step allowed for the anisotropic expansion of one side of the fabricated microbubbles by  nonuniform heating and associated anisotropic expansion. Our experimental setup consisted of two counter-propagating CO$_2$ laser beams (wavelength 10.6~$\mu$m) focused onto a fused silica capillary (250~$\mu$m and 350~$\mu$m as the inner and outer diameters, respectively). \begin{figure}[H]
    \centering
    \includegraphics[width=8cm]{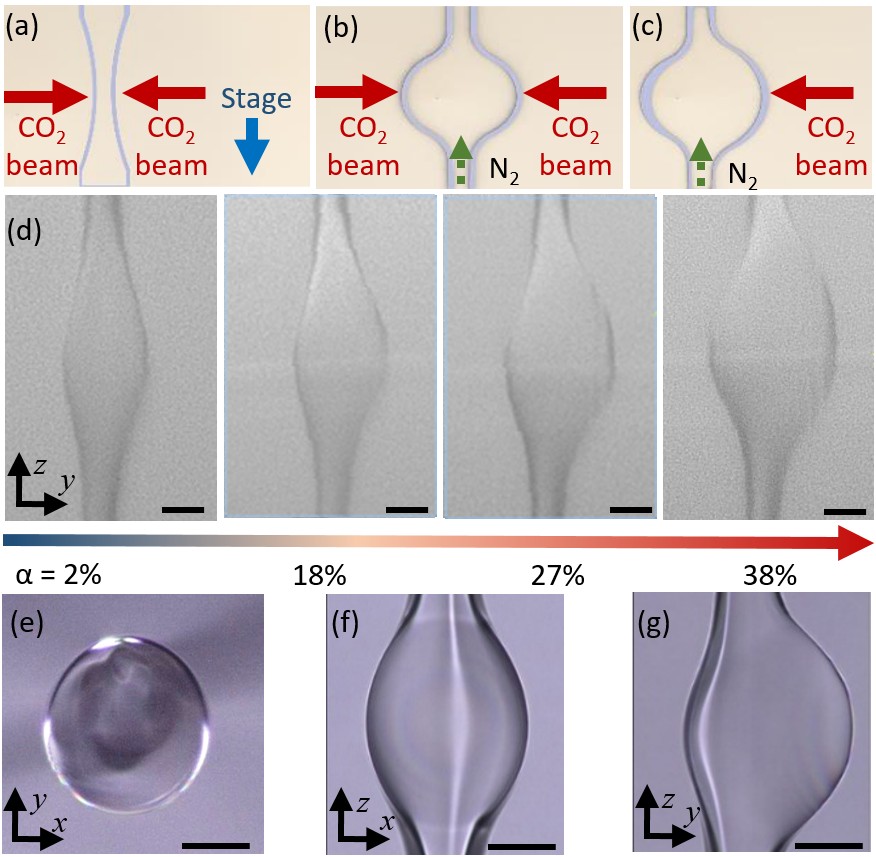}
        \caption{        Fabrication of a deformed, hollow resonator.  (a) A section of fused silica capillary, with inner and outer diameters of 250~$\mu$m and 350~$\mu$m, respectively, is tapered by shining a CO$_2$ laser on opposite sides while pulling via a single translation stage. (b) The softened tapered capillary is internally pressurized to isotropically expand, forming a microbubble.  (c) Asymmetry is introduced by continuing the expansion while using only a single CO$_2$ laser beam to illuminate the cavity. (d) Snapshots of the microbubble resonator during anisotropic expansion to an asymmetry percentage, $\alpha\approx38\%$. 
        Microscope images of a microbubble cavity after fabrication: (e) top view (\textit{xy}-plane), (f) side view (\textit{xz}-plane), and (g) side view (\textit{yz}-plane). Scale: 50~$\mu$m.} 
    \label{fab1}
\end{figure}
\vspace{1cm}
The bare silica capillary was first tapered by the heat-and-pull method using bidirectional illumination of the capillary via the CO$_2$ laser with a total power split between both beams of 5.5~W while being pulled on one end  via a translation stage, see Fig.~\ref{fab1}(a).   The tapered capillary was then pressurized using N$_2$ gas at 3~mbar while using a higher laser power (total power of $\sim7.5$~W) to expand the bubble isotropically to an outer diameter of $\sim$85~$\mu$m (Fig.~\ref{fab1}(b)).
The asymmetry was eventually introduced by blocking one of the counter-propagating CO$_2$ laser beams and further expanding the silica microbubble at a single beam power of $\sim$7.5~W - 12~W. This resulted in nonuniform softening of the silica wall and  anisotropic expansion of the pressurized microbubble (Fig.~\ref{fab1}(c)). 
The degree of asymmetry induced in the microbubble was controlled by the laser power and illumination time of the single CO$_2$ beam. Figure~\ref{fab1}(d) shows a sequence of frames of a microbubble being heated and expanded by a CO$_2$ beam incident from the right side. We introduce an asymmetry percentage parameter, $\alpha$,  calculated from microscope images, as the relative increase in microbubble size in the equatorial plane:
\begin{equation}
    \alpha = (R_{max} /R_{min} -1) \times 100,
    \label{alpha}
\end{equation}
where $R_{max}$ and $R_{min}$ denote the distances from the tapered capillary axis to the outer wall of the microbubble on its extruded (maximum distance) and non-extruded (minimum distance) sides, respectively. 
The microbubble appears as an egg-shaped silica shell in the \textit{xy}-plane view (Fig.~\ref{fab1}(e)) with an asymmetric profile in the \textit{yz}-plane (Fig.~\ref{fab1}(g)). The whispering gallery resonator (WGR) remains symmetric in the \textit{xz}-plane,  i.e., transverse to the CO$_2$ beam axis (Fig.~\ref{fab1}(f)).
\section{\label{Structchar}Structural Characterization}
X-ray tomography (Zeiss Xradia 510 Versa) was performed to determine the precise dimensions of the fabricated microbubbles \cite{chaos3d2025}, see  Appendix A. The deformed microbubble   resembles an asymmetric eggshell. An egg model \cite{egg} based on multipole deformations  fits the equatorial profile well:
\begin{equation}
    r(\theta) = r_0 \left(1 - \epsilon \sin\theta\right) \left(1 + \delta \cos^2\theta\right)
    \label{egg_eq}
\end{equation}
where the constants $\epsilon$ and $\delta$ determine the amplitudes of the vertically aligned dipole (limaçon-type) deformation and the $45^\circ$-tilted quadrupole deformation, respectively \cite{directionalemission_direction}. For one particular microbubble sample (see Appendix A), the observed asymmetry was 22$\%$. 
The third component of Eq.~\ref{egg_eq}, ($1 + \delta \cos^2\theta$), improves the geometric curve fitting by accounting for the smooth change in the radius, $r(\theta)$, from $R_{max}$ to $R_{min}$. Hence, $\delta$ is generally negligible since we consider a $\cos^2$ dependence on $\theta$, 
 implying that the equatorial profile of the microbubble is close to a limaçon shape, a result that was also confirmed for microbubbles with different asymmetry percentages up to 22$\%$.  Microbubble fragility made it difficult to obtain tomography data for larger asymmetries, limiting the range of estimates for the deformation parameter, $\epsilon$. However, the asymmetry percentage, $\alpha$, can be calculated directly from microscope images, as it depends only on the ratio of the radii, $R_{max}$ and $R_{min}$.   

It should be noted that, at the deformation levels considered in this work, the asymmetry percentage from Eq.~\ref{alpha} allows for the characterization of the dipole deformation of the microbubble. Taking the minimum and maximum radius values, $R_{min} = r_0(1-\epsilon)(1+\delta)$ and $R_{max} = r_0(1+\epsilon)(1+\delta)$, respectively, the asymmetry percentage is related to the contour fitting parameter, $\epsilon$, at negligible $\delta$ by 
\begin{equation}
    \alpha = 2\epsilon/(1-\epsilon) \times 100.
    \label{alpha_epsilon_relation}
\end{equation}
\section{Optical Characterization}
The asymmetric microbubble resonators were spectrally characterized using an optical microfiber-based near-field coupling scheme.  In all our measurements, the taper fiber was in contact coupling with the resonator, hence all reported Q-factors are loaded values.   The output of a wavelength-swept laser source (New Focus, Velocity TLB-6728) was coupled to an optical fiber circuit comprising a tapered fiber of $\sim 1$~$\mu$m in diameter, made from a 125~$\mu$m diameter single-mode fiber (Thorlabs, 1550BHP). The polarization of the guided light was adjusted via a three-paddle polarization controller. The fiber-transmitted signal, detected with an InGaAs photodiode (Thorlabs, PDB450C), was recorded and analyzed using a digital storage oscilloscope (Liquid Instruments, Moku:Lab).  

\begin{figure}[H]
    \centering
    \includegraphics[width=12cm]{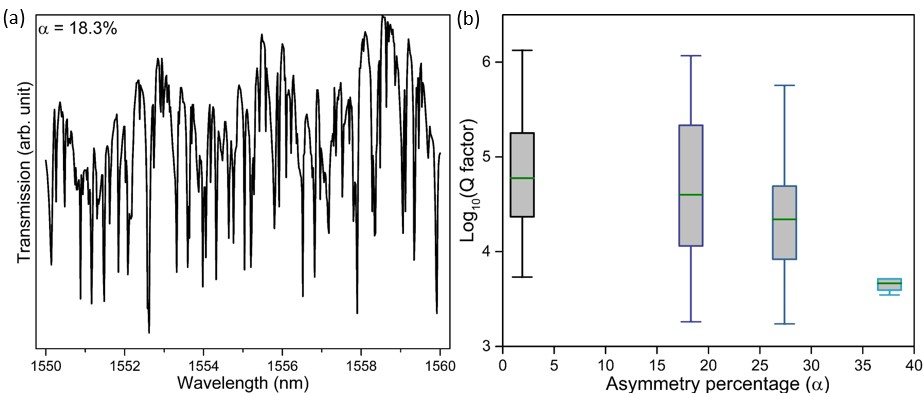}
    \caption{(a) Fiber transmission spectrum for a deformed microbubble with an asymmetry percentage, $\alpha = 18.3\%$.   (b) Loaded Q-factor distribution for deformed microbubbles with $\alpha$ varying from 1.9$\%$ to 37.6$\%$, where the initial undeformed microbubbles had near-identical diameters of 85~$\mu$m. The green line and gray box indicate the associated median and standard deviation, respectively. The whiskers extend from the minimum to the maximum loaded Q-factor observed experimentally across all modes within the wavelength scanning window of 1550~nm to 1560~nm.} 
    \label{Qdistribution}
\end{figure}

Figure~\ref{Qdistribution}(a) shows the transmission spectrum through the tapered fiber coupler recorded over a 1550-1560~nm wavelength range for a microbubble resonator with 18.3$\%$ asymmetry. The spectrum exhibits similar features to those observed in symmetric microbubble resonators \cite{Glasing2015}, i.e., a high density of high-Q-factor WGMs. The loaded Q-factor of each mode was measured individually from the recorded fiber transmission spectrum by taking the ratio of the central wavelength to the full width at half maximum of the associated mode, calculated using the Lorentzian peak-fitting function in Origin software.  
Most of the measured Q-factors are on the order of $10^4$ or $10^5$. A systematic analysis was conducted on four microbubbles with  $\alpha$ ranging from $ 1.9\%$ to $37.6\%$. The statistical results are displayed in Fig.~\ref{Qdistribution}(b). Despite a small decline in Q-factor, high-Q modes over $10^5$ were found for bubbles with asymmetry percentages up to 27.4$\%$. However, a severe drop was observed for higher asymmetry percentages. A loaded Q-factor up to $10^3$ was measured for a microbubble with 37.6$\%$ asymmetry.

\begin{figure}
    \centering
    \includegraphics[width=12cm]{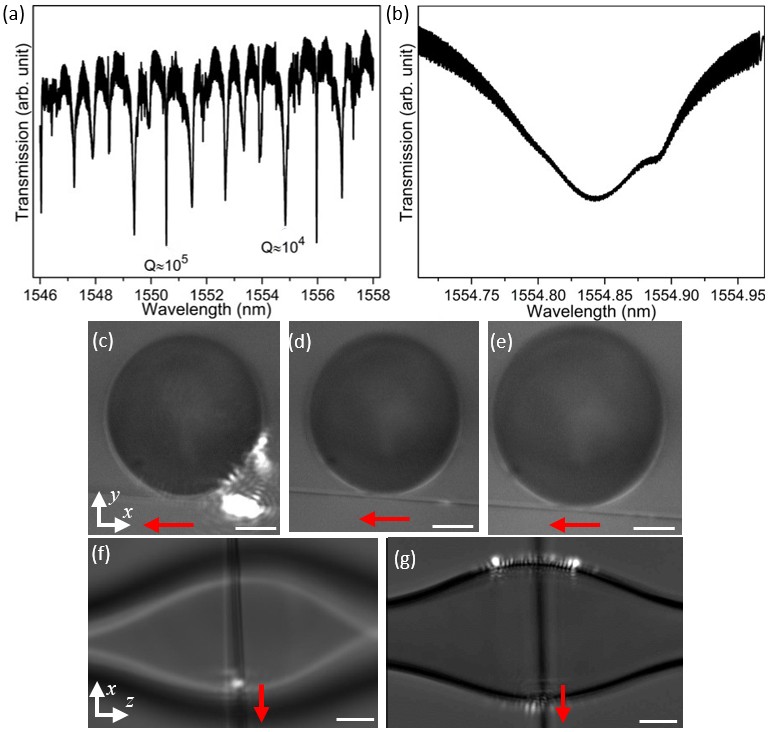}
         \caption{Optical characterization of a deformed microbubble resonator with an asymmetry percentage, $\alpha=15\%$, contact coupled to a fiber of diameter 1~$\mu$m at $\theta=0^\circ$. (a)~Fiber transmission spectrum. The fundamental mode and the emission mode are at 1550.54~nm and 1554.85~nm  with loaded $Q = 1.06 \times 10^5$ and $Q= 1.3\times 10^4$, respectively. (b)~High-resolution piezo scan of the emission mode with a deformed Lorentzian spectral shape. Near infrared imaging of the \textit{xy}-plane for (c)~the emission mode and (d)~the fundamental mode compared to (e)~off-resonance coupling.  
   Near infrared imaging of the \textit{xz}-plane for (f)~the fundamental mode and (g)~the emission mode with the number of intensity maxima associated with their axial orders, indicating that in this case the emission mode is a higher-order polar mode. Scale: 25~$\mu$m.} 
    \label{newanalysis}
\end{figure}

Spatial characterization of the emission from the WGMs was performed using a visible and short-wave infrared (SWIR) camera (Allied Vision, Goldeye G-130 TEC1) mounted on either of two orthogonal microscopes to observe top (\textit{xy}-plane) and side views (\textit{xz}-plane) of a deformed microbubble with $\alpha\approx15\%$. Figure \ref{newanalysis}(a) shows the transmission spectrum through the tapered fiber coupler recorded over a 1546-1558 nm wavelength range.
Each mode was then excited separately by continuously scanning the laser wavelength over a narrow spectral range via piezo control, then locking it to the detected resonance dip in the transmission signal.  For technical convenience, the tapered fiber was kept in contact at $\theta=0^\circ$ (according to Eq.~\ref{egg_eq}) with the deformed microbubble. 
Tuning the wavelength to a higher-order axial mode at 1554.85~nm, the resonance dip in the transmission spectrum distorts to a Lorentzian with a small kink, see Fig.~\ref{newanalysis}(b).
In this case, strong directional emission occurs from the thin-walled section of the asymmetric microbubble toward the input side of the optical nanofiber, as observed in Fig.~\ref{newanalysis}(c). The bright spot emanating from the nanofiber is caused by the strong scattering that occurs where the tapered fiber intersects the emission cone of the radiated light beam. 
The fundamental mode at 1550.54~nm is a non-emission mode, see Fig.~\ref{newanalysis}(d), with negligible radiation losses. The only visible difference with Fig.~\ref{newanalysis}(e), where no mode is excited (off-resonance coupling), is a weak light spot on the input side of the optical nanofiber.
The number of intensity maxima at the resonator surface, as visible in Figs.~\ref{newanalysis}(f) and (g), was used to identify the axial order of the mode.

\begin{figure}
    \centering
         \includegraphics[width=12cm]{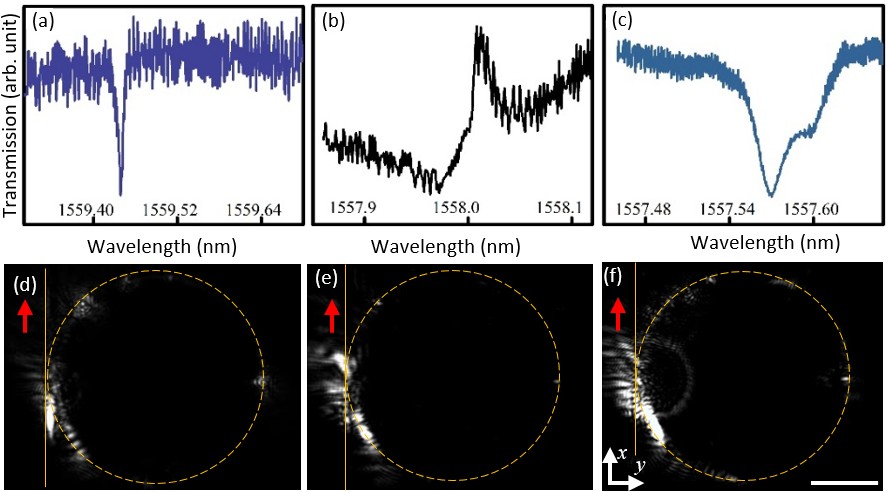}
     \caption{Spectral profiles of emission modes for a microbubble with asymmetry percentage $\alpha\approx12\%$ from (a) a Fano dip at 1559.44~nm with loaded $Q  = 2.1\times 10^5$, (b) a Fano peak at 1558.01~nm with loaded $Q  = 9.2\times 10^4$, and (c) a mode with a deformed Lorentzian shape at 1557.57~nm  with loaded $ Q  = 4.3\times 10^4$.  
     The associated emission profiles (d), (e), and (f), respectively, for a view in the \textit{xy}-plane. The red arrow indicates the direction of light propagation in the coupling fiber. Scale: 50~$\mu$m. }
    \label{obs}
    \end{figure}

\begin{figure}
    \centering
        \includegraphics[width=8cm]{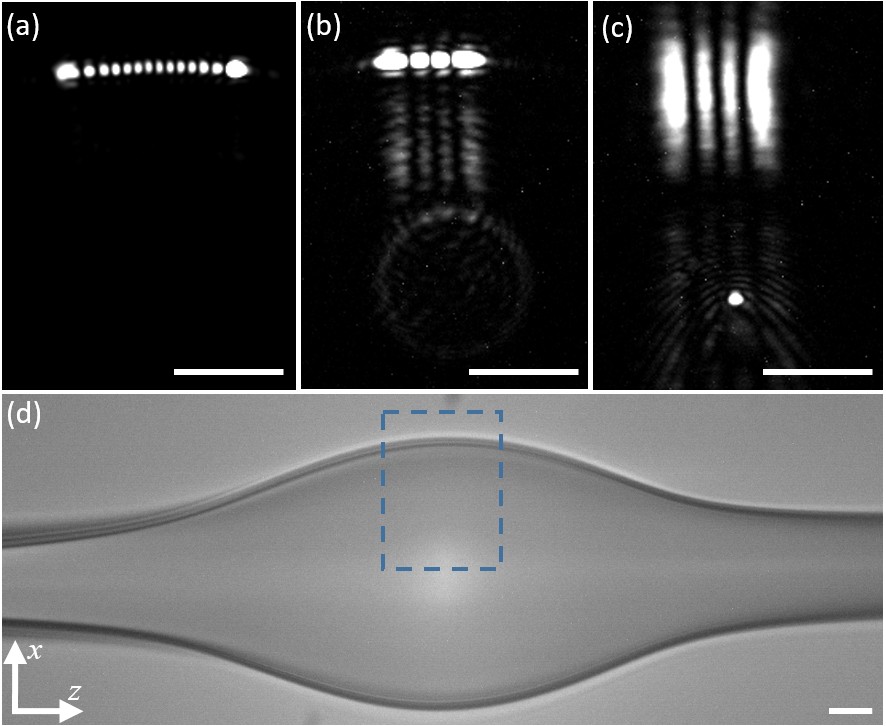}
    \caption{
    Near-infrared images of the \textit{xz}-plane of a  microbubble with an asymmetry percentage, $\alpha\approx12\%$, contact coupled to a  fiber of diameter 1~$\mu$m at $\theta=0^\circ$  (same as in Fig.~\ref{obs}(a)). The intensity distribution of (a) a non-emission mode at 1554.88~nm and an emission mode at 1559.44~nm, with a ring around the emission hotspot, when  the camera is focused on (b)  the top resonator surface and (c)  the emission hotspot. (d) Blue dashed square in the \textit{xz}-plane indicates region of interest for (a), (b), and (c). Scale: 20~$\mu$m.  }
    \label{S1}
\end{figure}

The emission properties were confirmed by repeating the experiment with a less asymmetric microbubble where $\alpha\approx12\%$. As shown in Fig.~\ref{obs}(a) to (c), this cavity exhibited more distinct resonance profiles for light-emitting modes. Corresponding \textit{xy}-plane images of the radiated light, where directional emission and the associated emission point can be seen, are shown in Fig.~\ref{obs}(d) to (f). Side-view pictures were also taken for the non-emission mode at 1554.88 nm, see Fig.~\ref{S1}(a), and the emission mode at 1559.44 nm, see Fig.~\ref{S1}(b) and (c). When adjusting the focus to the edge of the resonator, characteristic intensity distributions of higher-order polar modes were observed in both cases. However, an additional ring pattern appeared when the emission mode at 1559.44 nm was excited, see Fig.~\ref{S1}(b). Adjusting the focus to a different position on the resonator surface allowed us to identify a single, localized emission hotspot, visible in Fig.~\ref{S1}(c). It was verified that the presence of this emission hotspot was mode-dependent by sweeping the excitation wavelength over several resonances. The region of interest is marked in Fig.~\ref{S1}(d).


\section{ Simulation of Directional Emission of Higher-Order Radial Modes }
 
 Directional emission arises from symmetry breaking \cite{directionalemission2016}. In our microbubble resonators, two parameters contribute to breaking the rotational symmetry: the curvature of the outer interface and the wall thickness.  To gain more insight into the plausible physical mechanisms underlying our experimental observations, a two-dimensional (2D) COMSOL frequency-domain simulation was conducted to investigate the effect of asymmetry along the equatorial plane on directional emission. Although a 2D, rather than a three-dimensional (3D),  model was used due to limited computational resources, we assume that this provides a good approximation of  light propagation within the symmetry plane of deformed microbubble resonators.

A numerical model of a limaçon-shaped microring was constructed starting from a circular resonator of outer radius r$_0$~=~54~$\mu$m and wall thickness of 1.25~$\mu$m, with an additional dipolar deformation of amplitude $\epsilon$ applied to both the inner and outer interfaces. The value of $\epsilon$ was varied from 0.05 to 0.13 to vary the asymmetry of the resonator from 11$\%$ to 30$\%$. A 700~nm-wide bus waveguide positioned 500~nm away from the thicker part of the microring was used to excite the resonant modes and monitor the optical transmission. The transmission spectra predicted by the simulation model contain fewer modes than the experimental data, as only radial and azimuthal modes are considered in this 2D model. The numerical results show a decrease in Q-factors with increasing asymmetry, similar to the experimental trends in Fig.~\ref{Qdistribution}(b). The drop in Q-factor is significantly larger for higher-order radial modes as they experience a more drastic change in the propagation constant due to the varying wall thickness.

\begin{figure}
     \centering
     \includegraphics[width=8cm]{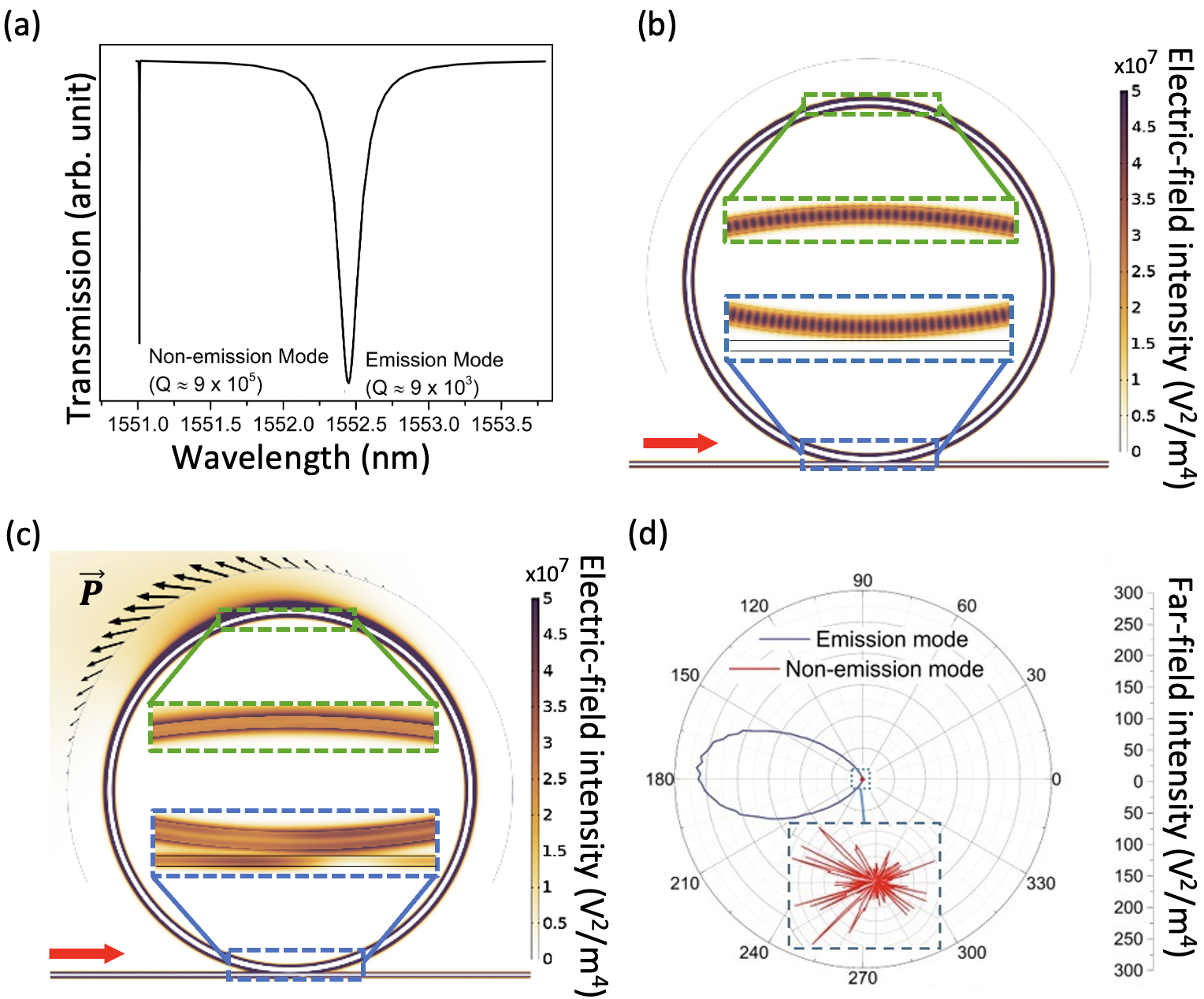}
     \caption{(a) The fiber transmission spectrum and simulated electric field profile in the \textit{xy}-plane of a deformed resonator with a dipole deformation amplitude, $\epsilon=0.13$, or an asymmetry percentage $\alpha\approx30\%$, coupled at $\theta=270^\circ$ for (b) a non-emission mode at 1551.011~nm with loaded $ Q=9\times 10^5$ and (c) an emission mode at 1552.45~nm  with loaded $Q=9\times 10^3$. Overlays in (b) and (c): zoomed-in images of the electric field distribution with arbitrary scale at $\theta=90^\circ$ (blue dashed box) and $\theta=270^\circ$ (green dashed box). The red and black arrows indicate the direction of propagation of light through the fiber and the Poynting vector, $\vec{P}$, respectively. (d) The  far-field profile of both the emission and the non-emission modes. Overlay: zoomed in area showing more clearly the far-field radiation pattern of the non-emission mode.}  
     
     \label{Simulation}
 \end{figure}

Figure~\ref{Simulation}(a) shows the simulated transmission spectrum obtained for the TM polarization in the 1551-1554 nm window when $\epsilon = 0.13$. Although the fundamental order radial mode has a loaded Q-factor of $9\times 10^5$, that of the first higher-order radial mode is two orders of magnitude lower, reaching only $9\times 10^3$. The corresponding spatial distributions of the external electric field intensity, i.e., the field outside the thin-walled resonator, are shown in Figs.~\ref{Simulation}(b) and (c), respectively. As expected, larger radiation losses are observed when the mode with a lower Q-factor is excited (Fig.~\ref{Simulation}(c)). However, those losses are highly anisotropic, as a significant amount of light is emitted from the thinnest region of the resonator. The black arrows show the Poynting vector distribution calculated along a circular arc surrounding the resonator. Owing to momentum conservation, the radiated light is mainly emitted in the direction tangent to the ring, and its orientation is fixed by the clockwise or counter-clockwise light propagation direction inside the resonator. As discussed in Appendix~B, similar numerical results were obtained after changing the orientation of the resonator and coupling the light at an angular position of $0^\circ$ (according to Eq.~\ref{egg_eq}) to reproduce the experimental conditions of the results shown in Figure \ref{newanalysis}.

The contrast between the two resonant modes can be explained by their different radial orders. The two overlays in Fig.~\ref{Simulation}(b) and (c) show the field intensity distributions in the thinnest (green dashed box) and thickest (blue dashed box) regions of the resonator, respectively. As shown in Fig.~\ref{Simulation}(b), the intensity of the fundamental mode remains well confined within the silica structure along the whole light path, including the thinnest section of the resonator. However, in the case of the first higher-order radial mode, the two intensity maxima of the mode profile shift from the field located within the silica structure in the coupling region (thickest part of the resonator) to the evanescent field located outside the silica structure in the thinnest section. This indicates that the waveguiding condition approaches, or possibly exceeds, the mode cutoff in the thinnest section of the resonator. The first higher-order radial mode is the only mode that becomes leaky in the thinnest part of the resonator and is therefore much more affected by the asymmetry of the resonator than the fundamental mode. For this reason, unidirectional emission occurs selectively when the first higher-order radial mode is excited.

A comparison between the far-field radiation patterns of the two modes is shown in Fig.~\ref{Simulation}(d). The radiation pattern of the first higher-order radial mode consists of a single emission lobe in the backward direction (close to $180^\circ$), which is consistent with the Poynting vector distribution calculated in Fig.~\ref{Simulation}(c). In contrast, the radiation pattern of the fundamental mode shows negligible far-field radiation at all angles, with no clear preferential emission direction. 

Both the field intensity distribution and the radiation pattern of the first higher-order radial mode differ from those reported  for solid limaçon-shaped 2D microresonators.\cite{directionalemission_direction,directional2009} In the case of the solid limaçon-shaped resonator, the directional emission governed by the chaotic nature of the modes occurs at different angular positions and in a direction orthogonal to the direction of the light emission observed in this work. Instead, in the case of the limaçon-shaped microring resonator, the localization of the emission near the thinnest part of the resonator and the tangential direction rather arise from the transition from a waveguided mode to a leaky mode over a restricted portion of the ring resonator.

\section{Discussion and Conclusion}
The goal of this work was to characterize the effect of asymmetry on microbubble WGRs. For this purpose, the conventional fabrication method for making microbubbles was modified to introduce controlled asymmetry. The reproducibility of the fabrication method is similar to that of conventional microbubbles, as the asymmetry is introduced by expanding with a single illumination beam  instead of two beams in the same setup. Similar to a microbubble, the desired dimension (or deformation) is achieved by gradually reducing the expansion rate at lower laser power and arresting the process once the dimension is reached.  
The geometry of the fabricated devices was characterized using X-ray tomography. The equatorial profiles of the microbubbles resemble those of an eggshell, the contour of which was fitted using a multipole deformation model inspired by limaçon and egg-shape curves.  We experimentally demonstrated that high Q WGMs still exist in microbubbles with a relatively large deformation and can be excited using an optical nanofiber-based light-coupling setup. Despite a slight decrease in average  Q-factors with increasing asymmetry, loaded Q-factors over $10^5$ were measured in deformed microbubbles with asymmetry percentages, $\alpha$, up to 27$\%$, equivalent to a dipole deformation amplitude, $\epsilon$, of $0.12$. Those results demonstrate that the radial light confinement in microbubble resonators makes them robust against structural asymmetry.

We also observed the directional emission of radiation losses for specific resonances with lower Q-factors. The majority of modes, especially those with higher Q-factors, do not show any significant sign of emission, and the mode-dependent emission hotspots are clearly distinguishable from the contact coupled fiber, confirming that the directional emission observed is not a consequence of scattering caused by the fiber or any particle. 
These directional emission properties may find applications in the design of WGM microlasers owing to the possibility of storing the gain medium within the WGR and the ultralow lasing threshold condition of high-Q WGMs. 
According to simulation results obtained from a 2D numerical model, a possible mechanism behind those directional emission properties could be the selective leakage of higher-order radial modes in the thinnest part of the microbubble. However, our simulation model only takes into account the change in wall thickness in the equatorial plane as measured from tomography data.
Changes in the axial confinement of the modes may also play an important role as our experimental observations relate to higher-order polar modes. A more realistic 3D model would be needed in order to gain deeper insight into a wider range of effects including polarization mixing, mode coupling, and mode leakage due to variations in the axial curvature and wall thickness of the resonator. Whether or not these effects could explain the experimentally observed emission hotspots or lead to the emergence of chaos in the optical response of deformed microbubbles remains an open question.

\section{Acknowledgments}
This research was supported by OIST Graduate University and JSPS KAKENHI Grant Numbers JP24K08290, JP23H04571, JP25K01643, JP25H01639. The authors would like to thank M.Z. Jalaludeen and S. Begumya for useful discussions, M. Ozer, K. Karlsson, and OIST's Scientific Imaging Section for technical assistance, and the Scientific Computing and Data Analysis (SCDA) section for providing the high-performance computing resources for COMSOL simulations.   

\subsection*{Appendix A: X-ray Tomography of Deformed Microbubble Resonators}
\begin{figure}[H]
    \centering
    \includegraphics[width=7cm]{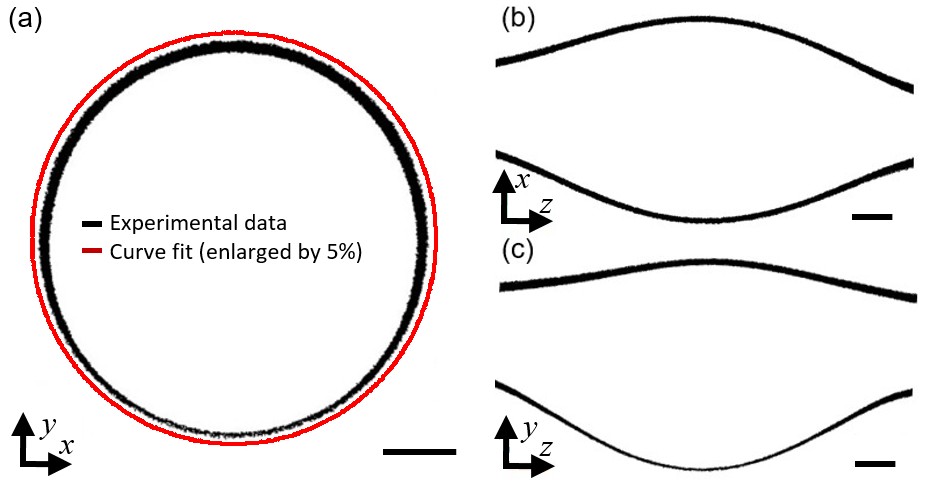}
    \caption{ X-ray tomography images of a sample microbubble processed by adjusting the threshold, removing the background, and inverting using ImageJ and its plugin. (a) Cross-section view in the \textit{xy} equatorial plane of a deformed microbubble with $\alpha=22\%$.  Overlay plot: outer wall curve fitted by Eq.~\ref{egg_eq} yielding $r_0=53.5~\mu$m, $\epsilon=0.1$, and $\delta=0.009$, enlarged by $5\%$ to distinguish it from experimental data. (b) View in the \textit{xz}-plane.  The symmetry profile is reasonably preserved.  (c)  View in the \textit{yz}-plane.  A significant extrusion is evident. Scale: $20~\mu$m. }
    \label{xray}
\end{figure}

The 3D datasets for X-ray tomography \cite{tomography2021} acquired for microbubbles with different asymmetry percentages were analyzed using ImageJ and the morpholibJ plugin \cite{morpholibj2016}. Figure~\ref{xray}(a) shows the equatorial cross-section of a microbubble with 22$\%$ asymmetry. As a result of the anisotropic expansion, the thickness of the silica wall gradually changes from 3.3~$\mu$m in the non-extruded region to  1.3~$\mu$m in the extruded region. The symmetric profile observed in the transverse \textit{xz}-plane  (Fig.~\ref{xray}(b)) is similar to that of a conventional microbubble \cite{Zia_wallthickness}. However, the profile observed in the \textit{yz}-plane (Fig.~\ref{xray}(c)) reveals a clear asymmetry in wall thickness between the extruded and non-extruded sides of the microbubble. The fit result indicates that, for this sample, $\epsilon=0.1$, $\delta =0.009$, and $r_0=53.5$~$\mu$m. Tracing back to the experimental value, $\alpha$, from the fit value, $\epsilon$, using Eq.~\ref{alpha_epsilon_relation} yields $\alpha = 22\%$.
   
\subsection*{Appendix B: Taper Fiber Coupling Condition}
\begin{figure}[H]
    \centering
    \includegraphics[width=8.cm]{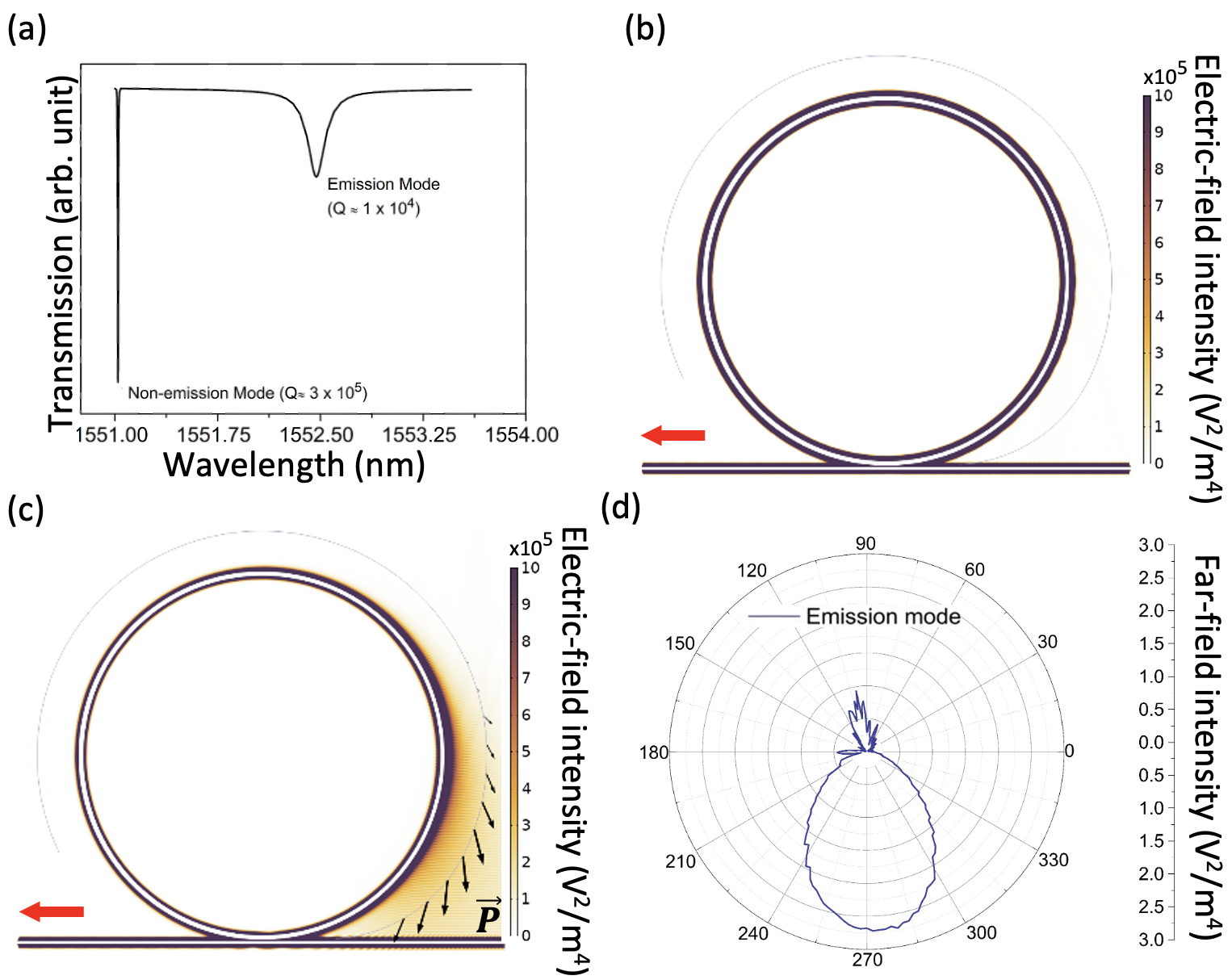}
    \caption{(a) The fiber transmission spectrum and simulated electric field profile in the \textit{xy}-plane of a deformed resonator with a dipole deformation amplitude, $\epsilon=0.13$, or an asymmetry percentage,  $\alpha\approx30\%$, coupled at $\theta=0^\circ$ for (b) a non-emission mode  at 1551.02~nm with loaded $Q=3\times 10^5$ and (c) an emission mode at 1552.47~nm with loaded $Q=1\times 10^4$.  The red and black arrows indicate the direction of propagation of light through the fiber and the Poynting vector, $\vec{P}$, respectively. (d) The far-field profile of the emission mode in (c).} 
    \label{fig7.2}
\end{figure}

The simulation presented in Fig.~\ref{Simulation}, in which the fiber was coupled at the thick-wall region of the microbubble at 270$^\circ$, was extended to include coupling at 0$^\circ$ to obtain the far-field profile, see Fig.~\ref{fig7.2}. 
Figure~\ref{fig7.2}(a) shows the simulated transmission spectrum obtained for TM polarization in the 1551-1554~nm window. While the fundamental order radial mode has a loaded Q-factor of $3\times10^5$, that of the first higher-order radial mode is relatively lower, reaching $1\times10^4$. The corresponding spatial distributions of the external electric field intensity (i.e. the field outside the silica material) are shown in Figs.~\ref{fig7.2}(b) and (c), respectively.
In this configuration, the Q-factor of the emission mode is relatively higher than for 270$^\circ$, and the emission from the deformed microbubble is directed toward the coupling fiber, see electric field profile and corresponding far-field profile in Figs.~\ref{fig7.2}(c) and (d), consistent with the experimental result in Fig.~\ref{newanalysis}(c).
\section*{Data Availability Statement}
The data are available from the authors upon reasonable request.
\section*{Conflicts of Interest}
The authors declare no conflicts of interest.
\nocite{*}
\nolinenumbers
\bibliographystyle{abbrv}
\bibliography{Egg_emission_references}
\end{document}